\documentclass{vgtc}                          

\input{config.sty}
\input{glossary.sty}

\usepackage{times}                     
\usepackage{mathptmx}

\onlineid{0}

\vgtccategory{Research}

\title{Constraint representation towards precise data-driven storytelling}

\author{Yu-Zhe Shi\thanks{e-mail: syz@autodsl.org}\\ %
    \parbox{1.4in}{\scriptsize \centering The Hong Kong University of \\ Science and Technology}
\and Haotian Li\thanks{e-mail: haotian.li@connect.ust.hk}\\ %
    \parbox{1.4in}{\scriptsize \centering The Hong Kong University of \\ Science and Technology}
\and Lecheng Ruan\thanks{e-mail: ruanlecheng@ucla.edu}\\ %
    \scriptsize Peking University %
\and Huamin Qu\thanks{e-mail: huamin@cse.ust.hk}\\ %
    \parbox{1.4in}{\scriptsize \centering The Hong Kong University of \\ Science and Technology}
}

\abstract{
    Data-driven storytelling serves as a crucial bridge for communicating ideas in a persuasive way. However, the manual creation of data stories is a multifaceted, labor-intensive, and case-specific effort, limiting their broader application. As a result, automating the creation of data stories has emerged as a significant research thrust. Despite advances in Artificial Intelligence, the systematic generation of data stories remains challenging due to their hybrid nature: they must frame a perspective based on a seed idea in a top-down manner, similar to traditional storytelling, while coherently grounding insights of given evidence in a bottom-up fashion, akin to data analysis. These dual requirements necessitate precise constraints on the permissible space of a data story. In this viewpoint, we propose integrating constraints into the data story generation process. Defined upon the hierarchies of interpretation and articulation, constraints shape both narrations and illustrations to align with seed ideas and contextualized evidence. We identify the taxonomy and required functionalities of these constraints. Although constraints can be heterogeneous and latent, we explore the potential to represent them in a computation-friendly fashion via Domain-Specific Languages. We believe that leveraging constraints will facilitate both artistic and scientific aspects of data story generation.
} 


\begin{document}

\maketitle

\section{Introduction}

Data-driven storytelling is a powerful vehicle for conveying ideas persuasively to target audiences. It has been widely applied in various scenarios~\cite{alharbi2021nanotilus,kouvril2021molecumentary,morth2022scrollyvis,chen2019designing}, including science education, clinical diagnosis with therapy interpretation, product popularization, and public policy advocacy, among others. Creating data stories is a comprehensive and costly effort that integrates multiple processes: mining relevant data, interpreting data insights, organizing textual narratives, and rendering visual illustrations~\cite{segel2010narrative,lee2015more}. Hence, the automatic generation of data stories is crucial for broader applications, despite its inherent complexity~\cite{wu2021ai4vis,li2023ai,li2024we}.

The major challenge of data-driven storytelling arises from the dual requirement of \emph{communicating subjective knowledge and insights} while \emph{supporting them with objective evidence}. When talking about a data story, we expect the insights behind the proposed therapy to align with the patterns entailed in clinical data regarding a healthcare propaganda; we expect the elements of the story to fairly reflect events happening in the real world regarding a theme-based news report; we expect to see photorealistic illustration animations with physically-real rendered ocean currents in an advocacy about marine pollution; and we expect to derive actionable messages that meet our realistic requirements when inquiring about government's record of export trade data. These \emph{twisted} expectations reveal the hybrid nature of data stories --- a blend of the imaginative aspects of storytelling and the grounded basis of evidence.

Data stories are not pure stories. A conventional story is constructed solely based on a core seed idea and is derived from the seed idea in a top-down manner. Stories make rationalization of scenarios, characters, and plots for a self-consistent virtual world with its unique dynamics~\cite{dubourg2022imaginary}. In contrast, data stories are required to be persuasive. Every piece of material used to organize the story should be grounded in evidence from the real world, thus limiting the extent of rationalization compared to purely fictional narratives. 

At the same time, data stories are not mere summarizations or reports of data. Those evidence-based approaches aim to reflect every aspect of data in detail, reconstructing the objective material without information loss in a bottom-up fashion. In contrast, data stories are required to convey specific, and sometimes opinionated, ideas rather than neutral ones~\cite{cetina1999epistemic}. All the evidence should compactly support the target idea, and the irrelevant information should be discarded, thereby refining the interpretation from a uniform one. 

Putting together, data-driven storytelling \emph{intertwines} the methodologies of top-down storytelling and bottom-up evidence-based analysis. If we consider a spectrum indicating the proportion of evidence and rationalization, data stories would lie in the middle of the two endpoints, mediating the properties of traditional stories and data reports. While a vast space of reasonable stories can be created through the lens of highly diversified perspectives, subjected to the authors' personal and societal contexts~\cite{heisenberg1958physics,bybee2006scientific,cetina1999epistemic,latour1986laboratory,latour1987science,lynch1993scientific}, there should exist a \emph{boundary} that restricts the space to a compact and permissible one, where the dual requirements of data stories, conveying ideas and grounding evidence, can be both satisfied.

This boundary can take various forms, such as domain-specific knowledge to ensure the integrity of data analysis, external memory of key variables to maintain logical coherence in the textual narrative, or models for model-based generative rendering to create physically-real visual illustrations. Regardless of its type, this boundary excludes the possibility of generating ambiguous or problematic data stories, thereby enhancing the precision of data story generation. The bounded permissible space, shaped by both seed ideas and the boundary, maximizes the flexibility of storytelling while ensuring that the story remains firmly grounded in evidence.

\begin{figure*}[t]
    \centering
    \includegraphics[width=\linewidth]{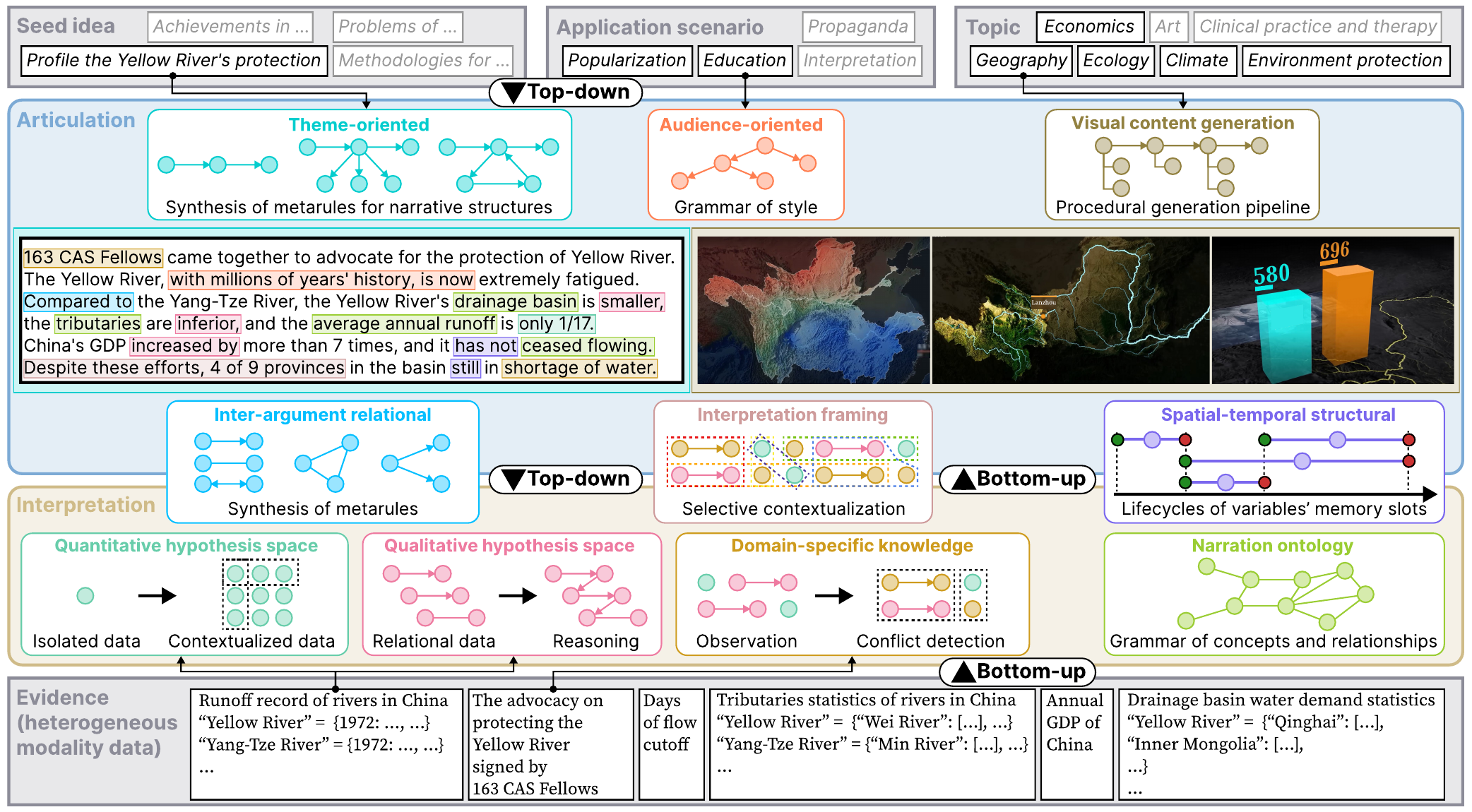}
    \caption{\textbf{The architecture of data-driven storytelling with hierarchical constraints.} We present intuitive illustrations of the representations with blocks (see \cref{subsec:dslascons}). The colors highlighting textual narratives and visual illustrations are encoded according to their respective constraints.}
    \label{fig:overview}
\end{figure*}

We refer to such boundaries as the \textbf{constraints} of data stories. Constraints, in contrast to production sets, determine whether a final product is \emph{feasible} under given requirements rather than defining the final product itself, which has been introduced to facilitate effective visualization creation, such as recommendation and verification~\cite{moritz2018formalizing,chen2021vizlinter,shen2021taskvis,yang2023draco,shankar2024spade}. In the context of data-driven storytelling, the production set is the seed idea of the story, which shapes the perspective to frame, the position to hold, and the context to include. Accordingly, we have various types of constraints at different hierarchies to ensure that data stories become what they should be.

Given the varied requirements of constraints, we hold the position that constraints are significant for generating data stories that are precisely coherent with both the author's intention and the grounded evidence. Unfortunately, although various generative tools have been developed, the efficient representation of constraints remains under-researched. This is not a trivial problem --- constraints are heterogeneous and often latent. As aforementioned, constraints can be first-order rules, higher-order rules, structural data abstraction, or even more specialized types according to domain knowledge. Additionally, constraints such as domain-specific knowledge and grammar for audience-oriented adaptation can be tacit knowledge held by domain experts~\cite{abend2008meaning}, making them difficult to be represented for computation~\cite{shi2022think}. With these challenges unaddressed, we suggest that the integration of constraints for data-driven storytelling still requires interdisciplinary research efforts.

In this perspective article, we first introduce the architecture of our cognitively inspired framework of constraint representation for data-driven storytelling (\cref{sec:arch}). Within this framework, we systematically study the requirements of constraints in different hierarchies, demonstrating with running examples (\cref{fig:overview,sec:cons4ds}). We also explore the potential of representing constraints with \acp{dsl} (\cref{tab:showcases}), along with the challenge and potential solutions for generalization across different domains and scenarios. Concluding with general discussions (\cref{sec:discuss}), we hope this paper provides the data-driven storytelling community with a fresh perspective on enhancing both \emph{the creativity of art} and \emph{the preciseness of science} in data story generation.

\section{Framework architecture}\label{sec:arch}

Our theoretical framework is grounded in the cognitive foundations of human perspective framing~\cite{shi2023perslearn}, aligning with the underlying logic of data-driven storytelling --- selectively integrating grounded supporting information to align with the central objective. Specifically, the constraints are defined across two hierarchies: \textbf{interpretation} and \textbf{articulation}. These hierarchies can be incorporated into two bidirectional pathways that characterize human information processing. The top-down view begins with a specific seed idea, moves to the connections between arguments and visualizations, and finally \textbf{interprets} individual pieces of evidence to support the arguments. Conversely, the bottom-up view starts with a piece of quantitative data insight, progresses to the qualitative logical relationships upon the insights, and ultimately returns to the narrations and illustrations that \textbf{articulate} these insights. 

\paragraph{Top-down}

There are several constraints to incorporate in the generation process from the seed idea to the arguments. We need a constraint with second-order rules to structure the narrative and illustrations, such as parallel arguments or progressive arguments~\cite{toulmin1958uses}. We also require a constraint with first-order rules to identify the relationships between arguments, such as contrast, enumeration, or analogy~\cite{crammond1998uses}. Additionally, a structural variable management mechanism is necessary to maintain logical coherence, such as managing a temporally varied key quantity in the data. Also, a grammar-based constraint is needed to tailor the narrative organization specifically for the target audience, as the same story should be told differently in contexts of scientific popularization versus therapy interpretation. Furthermore, the generation from arguments to evidence should also be constrained. While evidence itself cannot be ``generated'', the context of evidence can be carefully selected to support the arguments.

\paragraph{Bottom-up}

There are several constraints to incorporate in the extraction process from quantitative data insights to qualitative data insights. First, we need a constraint with flexible hypothesis spaces, equipped with composable functions to test arbitrary statistical relationships between arbitrary pairs or groups of variables, such as condition, correlation, and causality~\cite{ghahramani2015probabilistic}. Upon this layer, we need a constraint with combinatorial logic functions that abstract the quantities to relationships, such as opposites, comparisons, and superlatives. Additionally, we need a constraint that represents domain-specific knowledge to contextualize the data appropriately and detect potential insights. For example, in many cases, anomaly data can be insightful but can only be detected with an understanding of what constitutes normal data. Domain-specific knowledge also maintains the correctness of both quantitative and qualitative analyses, minimizing the possibility of misleading or deceptive interpretations. Furthermore, as the narration must be coherently aligned with the data of interest, the intersection between domain-specific knowledge and the scope of the selected data should also serve as a constraint on the grammar of the narration.

\section{Constraint representation}\label{sec:cons4ds}

In this section, we specify the taxonomies and functions of the constraints targeting precise data-driven storytelling. To ensure the descriptions are understandable, we first introduce a running example that goes throughout the remainder of this paper (\cref{subsec:cons4ds-eg}). Afterwards, we discuss the hierarchies to define the constraints (\cref{subsec:cons4ds-h}), and explore their representation approaches (\cref{subsec:dslascons}).

\subsection{An exemplar data story}\label{subsec:cons4ds-eg}

We carefully select a relatively ideal data story, \emph{The Protection of the Yellow River}\footnote{Please refer to the original video posted at \url{https://youtu.be/jq_74a58wtA?si=9LXODKpOipbmpWsS} (with English and Chinese captions)}, as our running example. The rationale for selecting this piece of work comes in three-fold: (i) all data insights are properly interpreted in the context of corresponding background knowledge; (ii) the narrative structure and style are adaptively optimized for both the topic itself and the target audience's expectations regarding a data story for education and advocacy; (iii) the visual illustrations upon the narrations are coherent and the photorealistic render visualizations are impressive. The story comes as follows.

\begin{quote}
    In 1997/98, 163 Chinese Academy of Sciences Fellows came together to advocate for the protection of Yellow River, when it failed to flow into the sea for nearly two-thirds of the year. At its worst, 90\% of the lower reaches of the river dried up completely. The Yellow River, with millions of years' history, is now extremely fatigued.

    Compared to the Yang-Tze River, the Yellow River's situation is even more dire. The drainage basin is significantly small, the tributaries are significantly inferior, and the average annual runoff is a mere 58B$m^3$, compared to the Yang-Tze River's 975.5B$m^3$, a ratio of just 1/17.

    A significant portion of the Yellow River's water, about 60\%, comes from just 28\% of the drainage basin above Lanzhou. The large water demand of 69.6B$m^3$ for industry further strain the river, against its 58B$m^3$ runoff.

    Consequently, the first flow cutoff occurred in 1972 and became frequent over the next two decades. By 1997, the flow cutoff problem had become increasingly severe, culminating in an unprecedented 226 days without flow. 

    [Measures have been implemented ...] From 1999 to 2022, China's GDP increased by more than 7 times, and the Yellow River has not ceased flowing. 
    
    However, the serious water and soil erosion of the Loess Plateau has made the Yellow River the river with the highest sediment content in the world.

    [Measures have been implemented ...] By 2022, compared to 1949, the vegetation coverage of the Loess Plateau had increased by 59\% to 65\%, and the amount of sediment entering the Yellow River had decreased by nearly 88\%, from 1.6B tons to 193M tons. 

    Despite these efforts, four of the nine provinces in the Yellow River drainage basin still suffer from extreme or severe water shortages, suggesting that the protection of the Yellow River still demands continuous effort.
\end{quote}

The data story is compressed due to space limits. We keep all narrations based on data insights and discard pure descriptive text for succinctness, \eg, the descriptions of measures and actions.

\subsection{The hierarchies of constraints}\label{subsec:cons4ds-h}

We define the constraints across the hierarchies of data-driven storytelling as the primary taxonomy. The hierarchical structure is intrinsically in line with the current consensus of the community~\cite{chevalier2018analysis}.

\subsubsection{Seed idea to articulation}

The seed idea is the starting point and the overall guidance of a data story. The seed idea reflects the internal factors, such as societal background, motivation, position, and perspective of the author regarding the topic~\cite{longino1990science}. For a high-quality data story, the narratives should be coherently subjected to the seed idea. Additionally, the narratives should also be adapted to external factors, such as the requirements of the presenter, the requirements of the target audience, the context of the presentation, and the objective of the presentation.

In our example, the seed idea is likely to be \emph{``To profile the past, the present, and the future of the Yellow River's protection''}, thus the narrative structure is organized as a sequence of \emph{problem-measure-result} triplets along the temporal dimension, within linear logic. If the seed idea changes to \emph{``The current problems of the Yellow River's protection''}, \emph{``The achievements in the Yellow River's protection''}, or \emph{``Methodologies for the Yellow River's protection''}, the narrative structure would vary accordingly. We refer to this constraint as the \textbf{theme-oriented constraint}.

Our exemplar data story is a video for popularization education, thus the narrative comes in a teaching style. If the presentation scenario changes to official propaganda or professional interpretation, the narrative style will vary accordingly. We refer to this constraint as the \textbf{audience-oriented constraint}.

As the topic covers disciplines such as geography, natural environment protection, climate, ecology, and economy, the visual illustrations are expected to be photorealistic. Specifically, data stories regarding geography and environment are also expected to visualize the data insights on 2D maps, 3D terrain models, or 3D globe models, and the visualization elements should come in a scientific style, \ie, the animations are physically real. For videos illustrating data stories of other topics, such as operational process instructions, historical event interpretations, or public policy advocacy videos, the languages of visual expression vary accordingly. We refer to this constraint as the \textbf{visual content generation constraint}.

\subsubsection{Articulation to interpretation}

Derived from and constrained by the seed idea, arguments are organized following a narrative structure. The arguments are not isolated at all --- instead, they are supported by the interpretations of evidence, and also, are then articulated in the global structure to fit with the context of each other~\cite{shi2023perslearn}. Similar to the former hierarchy, this hierarchy also only cares about the constraints derived from the high-level internal and external factors, rather than the data.

The connections between arguments specify the narrative structure. Different narrative structures determine distinct styles of articulation between arguments. In our example, the relationship \emph{``contrasting''} is a representative use case. To emphasize the severe condition the Yellow River was facing around 1997, by contrasting with the condition of the Yang-Tze River; and also, to highlight the achievements made by the protection actions, by contrasting key indicators before and after the treatments. We refer to this constraint as the \textbf{inter-argument relational constraint}.

Each argument must be supported by a piece of evidence. Argument and evidence are bridged by interpretations~\cite{kuhn1970structure}, which selectively construct different contexts as a premise, thus drawing distinct conclusions given the same set of observations. Most trickily, all aspects of the conclusions can be logically self-consistent. A typical case comes in the last sentence of our exemplar story --- given the same observation, which can be objectively described as \emph{``among the 9 provinces in the Yellow River drainage basin, 5 provinces are free from water shortage while 4 provinces are suffering extreme or severe water shortages''}. Interestingly, in the story, the context \emph{``despite these efforts, 4 of 9 provinces are suffering...''} modifies the audience's uniform prior on water shortage and constructs a context to increase their expectation of water shortage, thus enhancing the significance to the continuously protect the Yellow River. Through this frame of perspective, the interpretation of a piece of evidence becomes subjective, and it is logically correct. Let us consider another context, where we say \emph{``5 of 9 provinces are free from water shortages''}, the interpretation can be totally different --- we may be relieved that more than half of the provinces in the Yellow River drainage basin do not have the problem of water shortage. Thus, such interpretation is subjected to the arguments' roles, which is further determined by the author's perspective. We refer to this constraint as the \textbf{interpretation framing constraint}.

Along the development horizon of the story, some key concepts, patterns, and illustrations can be referenced by different arguments. They are thought to be invariant throughout the development process of the story, or they should be manipulated in a closure way, where each modification is coherent to its interaction with other concepts or patterns in the story. In our example, two distinct arguments call the concept \emph{``annual runoff of the Yellow River''}, which are required to keep in mind that these two concepts are identical without any variants, sharing the same quantity. We refer to this constraint as the \textbf{spatial-temporal structural constraint}.

\subsubsection{Evidence to interpretation}

Interpretation from evidence is the foundation of a data story. In contrast with the former two hierarchies, which are defined from the top-down view, this hierarchy is defined from the bottom-up view, where we are detecting insights from the data, and only from the data, regardless of the high-level factors.

Elementary data insights are entailed in quantities. However, the absolute quantities of individual variables may not be insightful. Insights come from the combinations of and the associations between different variables. In our example, the fact that \emph{``the average annual runoff of the Yellow River is 58B$m^3$''} is not insightful at all, since it is sufficiently a large number for ordinary people. However, contextualizing this piece of data together with \emph{``the average annual runoff of the Yang-Tze River is 975.5B$m^3$''} makes the analysis insightful --- although people cannot make precise perception of the absolute quantities of 58B and 975.5B, \ie, ``how large they are'', the distinction between the two distributions is trivial --- significantly, the former is much smaller than the latter, conveying the idea of the Yellow River's inferior condition to the audience. Such insights come from the appropriate construction of hypothesis spaces over statistics variables, which are subjected to the feasible combinatorial spaces of individual variables. We refer to this constraint as the \textbf{quantitative hypothesis space constraint}.

Logical insights are built upon the data insights and come from reasoning over logical relationships between variables for reasoning, which take a further step from evidence to interpretation. In our example, \emph{``the annual water demand for industry is 69.6B$m^3$, larger than the annual runoff of the Yellow River''} entails a logic proposition \emph{``Industry is extremely water demanding''}. Given another proposition \emph{``Beijing-Tianjin-Hebei metropolitan area is a major industry area of China and relies on the Yellow River for water''} and the general commonsense \emph{``Industry is the major driving force of GDP''}, we can easily find the outcome \emph{``China's GDP increased by more than 7 times while the Yellow River has not ceased flowing''} quite interesting and persuasive, demonstrating the effect of practical measurements such as optimizing water allocation policy. These insights come from multiple-step logical deduction, induction, and abduction, which are subjected to feasible hypothesis spaces spanned by logic operators. We refer to this constraint as the \textbf{qualitative hypothesis space constraint}.

In addition to those insights that can be detected by contextualization, another family of insights seems not insightful even in the context of associated variables or logic chains. They usually come from \emph{anomaly}. Although seeming normal, those data or logic may become abnormal when contextualized in domain-specific knowledge, which is sometimes not as trivial as general commonsense. In our example, \emph{``60\% water of the Yellow River comes from just 28\% of the drainage basin above Lanzhou''} is insightful only given the background knowledge of the positive correlation between water runoff and the area of drainage basin. Similarly, in \emph{``163 \ac{cas} Fellows came together to advocate for the protection''}, the quantity 163 will seem insightful only with the background knowledge that there were about 300 \ac{cas} Fellows in total by 1998. Such insights, coming without intrinsic statistical or logical patterns of interest themselves, can only be detected under the constraint of background knowledge. We refer to this constraint as the \textbf{domain-specific knowledge constraint}.

These constraints for interpretation we have analyzed will be integrated into the narrations ultimately. Thus, along with the high-level factors, the ontology extracted from the data also constrains the generation of narrations. Specifically, ontology defines the terminologies and the relationships between them, according to the domain reflecting the data. Our example comes with the domain-specific terminologies, such as \emph{``drainage basin area''}, \emph{``the number and length of tributaries''}, \emph{``average annual runoff''}, and \emph{``days of flow cutoff''}, and also the relationships between them, such as \emph{``the significant positive correlation between the scale of tributaries and average annual runoff''}, \emph{``the weak negative correlation between average annual runoff and days of flow cutoff''}, and \emph{``a tributary is counted in the drainage basin of its mainstream''}. This ontology is then integrated into the grammar of narratives to ensure that there are neither missing nor redundant terminologies and relationships regarding the source domain. Namely, the grammar thereby \emph{compactly} tailors the knowledge and insights behind the evidence. We refer to this constraint as the \textbf{narration ontology constraint}.

\begin{table*}[t!]
    \centering 
    \caption{\textbf{Demonstrations of constraint representations}}
    \rowcolors{1}{gray!20}{white}
    \renewcommand{\arraystretch}{1.25}
    \begin{tabularx}{\linewidth}{c|X|>{\arraybackslash}X}
    \toprule
    \textbf{Constraint} & \textbf{Showcase in our exemplar}  & \textbf{\ac{dsl}-based constraint representation}\\
    \midrule 
    Theme-oriented & \emph{The motivational seed idea of the data story is to profile the past, the present, and the future of the Yellow River's protection.} & \cdfont{\cdfun{time\_linear}(\cdfun{P}, \cdfun{Q}, \cdfun{R}) :- \cdkey{past}(\cdfun{P}), \cdkey{present}(\cdfun{Q}), \cdkey{future}(\cdfun{R}).\newline \cdfun{protection}(\cdfun{X}, \cdfun{Y}, \cdfun{Z}) :- \cdfun{problem}(\cdfun{X}), \cdfun{measure}(\cdfun{Y}), \cdfun{result}(\cdfun{Z}).\newline \cdkey{nested}(\cdfun{time\_linear}, \cdfun{protection}).} \\
    Audience-oriented & \emph{The data story comes for popularization education, thus the narrative should come in a teaching style, targeting for specialized audience group.} & \cdfont{\cdfun{teaching\_narrative} ::= \cdfun{engage} \cdkey{|} \cdfun{explore} \cdkey{|} \cdfun{explain} \cdkey{|} ... \newline \cdfun{engage} ::= \cdstr{"Let us"} \cdkey{|} ... \newline ...}\\
    Visual content generation & \emph{The topic is about geography, natural environment protection, climate, ecology, and economy, the visual illustration is expected to be photorealistic.} & \cdfont{\cdstr{3D\_terrain} = \cdfun{SurfaceModeling}(\newline\cdfun{size} = [\cdfun{L}: \cdcon{100}, \cdfun{W}: \cdcon{100}, \cdfun{H}: \cdcon{50}], \newline\cdfun{camera} = [\cdfun{X}: \cdcon{20}, \cdfun{Y}: \cdcon{30}, \cdfun{Z}: \cdcon{45}], \newline\cdfun{base\_model} = \cdstr{"YellowRiver.asm"}, ...)} \\
    Inter-argument relational & \emph{Compared to the Yang-Tze River, the Yellow River's drainage basin is significantly smaller, the tributaries are significantly inferior, and the average annual runoff is significantly lower.} & \cdfont{\cdfun{YR\_vs\_YT}(\cdfun{X}, \cdfun{Y}) :- \cdkey{duplicate}(\cdkey{contrast}(\cdfun{X}, \cdfun{Y})).\newline \cdfun{YR\_vs\_YT}(\cdstr{YR}, \cdstr{YT}) :- \cdfun{smaller\_basin}(\cdstr{YR}, \cdstr{YT}), \cdfun{inferior\_tributary}(\cdstr{YR}, \cdstr{YT}), \cdfun{lower\_runoff}(\cdstr{YR}, \cdstr{YT}).} \\
    Interpretation framing & \emph{Despite these efforts, four of the nine provinces in the Yellow River drainage basin still suffer from extreme or severe water shortages.} & \cdfont{\cdfun{water\_supplying} ::= \cdstr{sufficient} \cdkey{|} \cdstr{shortage} \cdkey{|} \cdstr{severe\_shortage} \newline \cdkey{interpret}(\cdfun{water\_supplying} -> \cdkey{!}\cdstr{sufficient}).}  \\
    Spatial-temporal structural & \emph{The average annual runoff is a mere 58B$m^3$... The large water demand for industry further strain the river, against its 58B$m^3$ runoff.} & \cdfont{\cdstr{YR\_runoff} = \cdkey{new memory slot} \cdfun{X}. \newline ... \newline \cdstr{YR\_industry} = \cdkey{new memory slot} \cdfun{Y}. \newline \cdkey{cmp}(\cdstr{YR\_runoff}, \cdstr{YR\_industry}) :- \cdkey{cmp}(\cdkey{call}(\cdfun{X}), \cdkey{call}(\cdfun{Y})).}\\
    Quantitative hypothesis space & \emph{The average annual runoff is a mere 58B$m^3$, compared to the Yang-Tze River's 975.5B$m^3$, a ratio of just 1/17.} & \cdfont{\cdkey{dyadic\_relation}(\cdfun{X}, \cdfun{Y}) -> \cdkey{stat\_prop}(\cdfun{X}, \cdfun{Y}). \newline \cdstr{YR\_runoff} = \cdcon{[]}, \cdstr{YT\_runoff} = \cdcon{[]}.\newline \cdkey{stat\_prop}(\cdstr{YR\_runoff}, \cdstr{YT\_runoff}).} \\
    Qualitative hypothesis space & \emph{From 1999 to 2022, China's GDP increased by more than 7 times, and the Yellow River has not ceased flowing.} & \cdfont{\cdkey{corr}(\cdstr{GDP}, \cdstr{YR\_industry}). \newline \cdkey{corr}(\cdstr{YR\_industry}, \cdstr{YR\_cutoff}). \newline \cdkey{map}([\cdstr{GDP}.\cdkey{pre}, \cdstr{GDP}.\cdkey{post}] -> [\cdstr{YR\_cutoff}.\cdkey{pre}, \cdstr{YR\_cutoff}.\cdkey{post}]).} \\
    Domain-specific knowledge & \emph{A significant portion of the Yellow River's water, about 60\%, comes from just 28\% of the drainage basin above Lanzhou. \newline In 1997/98, 163 Chinese Academy of Sciences Fellows came together to advocate for the protection of Yellow River.} & \cdfont{\cdkey{corr}(\cdstr{water\_pc}, \cdstr{basin\_pc}).\newline \cdstr{water\_pc} = \cdcon{60}, \cdstr{basin\_pc} = \cdcon{28}. \newline \cdfun{abnormal}(\cdkey{corr}(\cdstr{water\_pc}, \cdstr{basin\_pc})). \newline \cdstr{num\_Fellow} = \cdcon{300}. \newline \cdfun{abnormal}(\cdkey{!significance}(\cdstr{num\_Fellow}, \cdcon{163})).}\\
    Narration ontology & \emph{The terminologies and the relationships between them, according to the domain reflecting the data, should be integrated into the narratives ultimately.} & \cdfont{\cdkey{terminology} ::= \cdfun{drainage\_basin} \cdkey{|} \cdstr{basin\_area} \cdkey{|} \cdstr{annual\_runoff} \cdkey{|} ... \newline \cdkey{relationship} ::= \cdkey{corr}(\cdstr{basin\_area}, \cdstr{annual\_runoff}) \cdkey{|} \cdfun{in\_drainage\_basin\_of}(\cdstr{tributary}, \cdstr{mainstream}) \cdkey{|} ...}\\
    \bottomrule
    \end{tabularx}
    \label{tab:showcases}
\end{table*}

\subsection{DSL as constraint representation}\label{subsec:dslascons}

Given the taxonomy of the constraints \wrt their requirements, we explore how to \emph{represent} these constraints. Summarizing the properties and requirements of the ten constraints, we suggest that structural representations tailored for the constraints' definitions, namely \acp{dsl}, may become the appropriate approaches. We also showcase the utilities of \ac{dsl}-based constraints in our exemplar (\cref{tab:showcases}).

\subsubsection{The rationale of DSL-based constraints}

The constraints we have discussed indeed share some commonalities: they are required to be complete when verifying the generated content, avoiding open-ended cases; they must be consistent as the story progresses, avoiding ``magical modifications''; they are also expected to precisely encode knowledge of various granularity. These properties --- completeness, consistency, and multiple-granularity --- naturally fit the advantages of symbolic representation, in particular through programming languages, the symbolic representation with the highest expressive capacity~\cite{tarski1946introduction,chomsky1957syntactic,hopcroft1996introduction,russell2010artificial}.

Unfortunately, a major part of programming languages, the \acp{gpl} such as C/C++, Python, and Java, may not be the best candidates to represent the constraints --- programs written in those languages can become extremely complicated, thus hindering both machine program generation and human understanding~\cite{mernik2005and,fowler2010domain}. As \acp{gpl} maintain a general set of syntactic and semantic features to cover all aspects of usage, \ac{gpl} programs for a narrower set of usage are also built from those features from a relatively low level of abstraction. In contrast, \acp{dsl} that consider a specific set of usage only introduce features tailored for the target domain, thus enjoying a higher level of abstraction. This results in simple programs only with features echoing the domain-specific requirements, such as domain knowledge, which are easy to synthesize by machines, and are also easy to learn, understand, and use by domain experts without programming experience. 

The constraints for data-driven storytelling are heterogeneous, such as representing structures, knowledge, models, and calculations, respectively. In addition, some of them are tacit knowledge of domain experts, which requires fine-grained domain-specific knowledge injection. Consequently, they are appropriate to be represented with \acp{dsl} --- one \ac{dsl} for one specific type and one specific domain. Such considerations are generally acknowledged. For example, there is a variety of \acp{dsl} developed for creating diverse visualizations targeting specific domains efficiently~\cite{mcnutt2022no}.

In the following paragraphs, we discuss the utilities of \ac{dsl}-based constraints, according to the abstraction levels they are working on~\cite{abelson1996structure}: (i) \textbf{syntax-level constraints} care about the structures of structural representations, such as trees, graphs, and cycles, and also the mechanisms of symbolic calculation, such as unary, dyadic, and multiple operators; (ii) \textbf{semantics-level constraints} consider the exact meanings of variables, operators, and functions, echoing the ontology of the reference model from the source domain; (iii) \textbf{execution-level constraints} synthesize and interpret programs dynamically, namely linking and contextualizing unit components in the programs and verifying their global consistency. 

\subsubsection{Syntax-level constraint}

Among the ten constraints, theme-oriented constraint and inter-argument relational constraint are working on the syntax level. Their major utilities are generating meta-level templates, \aka metarules~\cite{emde1983discovery}, for defining a feasible space and permissible operations upon the space. Afterwards, the narrations are generated by grounding the space without conflicts with the constraint. 

Theme-oriented constraint shapes the narration structures, which are usually tree-based or graph-based. For different seed ideas, we may exploit different narrative structures to maximize their communication bandwidths. According to the theories of arguments~\cite{toulmin1958uses}, we may use linear structure for temporal-related contents, multi-headed structure for spatial-related contents, and non-monotonic logical structure for contents with subjective judgments. We can also locally nest different types of logic for mixed purposes. Similarly, inter-argument relational constraint implements the narration structures. There are sequences for progressive arguments, branches for alternative arguments, recursions for repeatedly updating arguments, and parallels for contrasting arguments.

\subsubsection{Semantics-level constraint}

Among the ten constraints, audience-oriented constraint, visual content generation constraint, interpretation framing constraint, and narration ontology constraint are working on the semantics level. Their major utilities are ensuring the specific meanings of the generated content to be consistent with general commonsense and domain-specific knowledge, and also to be complete for use.

Audience-oriented constraint and narration ontology constraint both shape the generation space of textual narrations. The former comes from a higher level, \ie, external factors of the data story, while the latter comes from a lower level, \ie, the given evidence. These two constraints are usually represented as deterministic or probabilistic \acp{cfg}, controlling the style and scope of the narratives~\cite{hopcroft1996introduction}. To constrain the style, we leverage the combination rules of specialized keywords, sentence structures, and transitions between sentences. For example, we use engaging transitions like \emph{``... Now it is the turn to do it together ...''} in data stories for education; we exploit keywords with sense-of-belonging, such as \emph{``our community''} in data stories for advocacy; and we employ sentences with superlative statements, such as \emph{``... is the highest/ best of ...''} in data stories for propaganda. To constrain the scope, we map the ontology from the corresponding domain of the evidence to the abstract grammar of narratives, both completing the concepts inside the scope and removing those out of the scope.  

Interpretation framing constraint can be viewed as a probabilistic \ac{cfg}, which is a tree with intermediate nodes spanning the \emph{world}, \ie, all possible candidate meanings, of specific concepts or events~\cite{reiter1981closed}. The world is the context for interpretation, mostly coming from the general commonsense. The process of framing is reweighing the candidates belonging to the same world.

Visual content generation constraint is the \emph{model} for model-based generation. For data stories on topics related to natural sciences, clinical practices, and engineering, visual illustrations are often required to be physically real. Despite the current advancements of \ac{aigc} techniques, generating photorealistic videos that are physically real is still challenging because elementary physical properties, such as the spatial-temporal dynamics, are latent and long-tail distributed in datasets, implying that they may not be correctly extracted during training. Instead, they may be induced as shortcuts. This is the drawback of model-free generation by nature. Consequently, we may leverage the physical constraints provided by \acp{dsl} for 3D modeling, such as Blender\footnote{\url{https://www.blender.org/}} --- we can synthesize Blender code for programs rendering a 3D model, precisely edit the model by modifying the program, and explicitly constraint the model with physical properties from \emph{the first principle}. Furthermore, for data stories on topics related to history, public policy, and business, visual illustrations are usually expected to be animated drawings rather than photorealistic videos. However, those animations can be a sophisticated combination of components, such as spatially articulated objects and temporally varied scenes, where the similar-sample-based end-to-end generative models may be struggling~\cite{shi2023complexity}. The straightforward solution also leverages a rule-based model generating local states, layout and rendering configurations, topological relationships, and temporal state transitions of the components.

\subsubsection{Execution-level constraint}

Among the ten constraints, spatial-temporal structural constraint, quantitative hypothesis space constraint, qualitative hypothesis space constraint, and domain-specific knowledge constraint are working on the execution level. Their major utilities are generating hypothesis spaces dynamically and verifying them in real-time. 

Quantitative and qualitative hypothesis space constraints are dynamically generating hypothesis spaces to detect any possible data insights, \ie, evidence of interest. A piece of interesting evidence with insight comes from the shift from one way of explanation to another, akin to the moment of representation shift in problem-solving~\cite{auble1979effort,kounios2009aha,ohlsson1984restructuring}. Analogous to the representation of a problem, which determines \emph{selecting what information of the problem into solving it}, our hypothesis spaces consider \emph{putting which pieces of evidence together to explain them}. For example, the data on \emph{``the Yellow River's annual runoff in 1998''} possesses multiple contexts, such as the 1998 annual runoff of other rivers in China, the Yellow River's annual runoff in other years, the industry water demand of the Yellow River drainage basin in 1998, and the annual runoff of the upper Lanzhou part of the Yellow River. The hypothesis space indicates the structure of observable variables, \eg, dyadic or triadic, and the type of verification, \eg, statistical testing functions or logical reasoning functions. Thereby insights can be detected at the shifting from isolated data to contextualized data. Domain-specific knowledge constraint works with quantitative and qualitative hypothesis space constraints. While the latter two put evidence together in different frames, the former puts grounded background knowledge, either procedural or declarative knowledge, together with evidence, to create insightful context shifting. 

Spatial-temporal structural constraint is a robust infrastructure for demonstrating the detected insights in the narrations. The variables are called in distinct parts and by various means, necessitating the maintenance of numerical integrity and logical consistency. On the temporal dimension, the lifecycles of the invariant are tracked to avoid inconsistency between different calls. Also, variables being modified with specific calculations are constrained with preconditions and postconditions for state transition tracing. On the spatial dimension, the relative changes of variables are tracked, such as duplication of variables, chaining among triple variables, inversion between dual variables, and recursion on multiple variables.

\section{General discussions}\label{sec:discuss}

In this perspective, we propose integrating constraints into the automatic generation of data stories to facilitate the creativity in storytelling alongside the preciseness in data analysis. We investigate the requirements of these constraints and explore the possibility of representing them through \acp{dsl} based on a realistic example. It is important to note that the proposed taxonomy of constraints may not be entirely mutually exclusive and collectively exhaustive --- there may be other specific constraints that are significant in different data stories and cannot be perfectly categorized within our definition of constraints. Our primary aim is to provide a structured framework for the data-driven storytelling community, which may, in turn, inspire the development of a fine-grained taxonomy of constraints and their corresponding implementation techniques.

\subsection{Integrating constraints into the current workflow}

The data-driven storytelling community has made significant efforts in developing powerful tools for creating data stories. Currently, there are two schools of thought on the generative models --- multi-stage pipelines and end-to-end approaches. We propose that constraints should be incorporated into both approaches, through implicit and explicit methods, respectively. In multi-stage pipelines, different categories of constraints can be mapped to corresponding stages~\cite{li2024we} --- such as analysis~\cite{wang2019datashot}, planning~\cite{zhao2021chartstory}, implementation~\cite{tyagi2022infographics}, and communication~\cite{hall2022augmented} --- and the modules within these pipelines can be modified according to these constraints. Additionally, the generated content of different modules can be verified against the relevant constraints, thereby implicitly integrating constraints into the workflow. For end-to-end approaches, which feature a higher degree of integration across the entire workflow~\cite{lu2021automatic}, constraints can be explicitly added by appending a constraint layer to the final output of the tools and verifying the generated content against these constraints. In this way, we outline a framework for integrating constraints into the generation workflow, which may inspire further research on refining the individual tools within the workflow with constraints at a more granular level.

\subsection{Automating the entire workflow with constraints}

Our ultimate goal is to automate the entire workflow of generating data stories, necessitating the automatic synthesis of constraints rather than manual specification. Code generation techniques facilitate the automatic synthesis of constraint programs~\cite{zan2023large,liang2024large}, given simple instructions, requirements, or textual narrations to be verified. Subsequently, the satisfaction of these constraints is verified through language features, such as answer set planning over logic programs~\cite{lifschitz1999answer}, which has been applied to constrain visualizations with design theories~\cite{moritz2018formalizing}. Although this is a rudimentary approach to utilizing constraints, it treats constraint program verification as \emph{``first-class citizens''}, ensuring determinism. All uncertainties, ambiguities, and factual errors introduced by generative models with non-deterministic nature, such as the hallucination of \acp{llm}, are subject to the top-level constraints. 

Consequently, this framework preserves inherent freedom of the random creativity characteristic of \ac{aigc}, while simultaneously ensuring that this creativity operates within a secure environment. We hope this straightforward yet self-consistent framework can serve as an accessible starting point for exploring the synthesis of constraints through active interaction with generative models. Indeed, the framework includes objective, subjective, and context-dependent constraints. Quantitative and qualitative hypothesis space constraints, spatial-temporal structural constraint, and visual content generation constraint are exactly objetive constraints, reflecting data and the physical world. In contrast, theme-oriented constraint, audience-oriented constraint, and interpretation framing constraint are relatively subjective, influenced by human preference. Additionally, domain-specific knowledge constraint and narration ontology constraint are context-dependent, sensitive to the exact scope of the evidence and the target story. We would like to clarify that the implementation of those heterogeneous constraints according to different requirements, such as transforming domain-specific knowledge into corresponding constraints, is beyond the scope of this perspective article. Nonetheless, explicitly disentangling objective, subjective, and context-dependent constraints within our proposed framework and exploring their respective implementations represents a significant direction for future research.

\subsection{On the generality of constraints}

While it is theoretically possible to automate the entire workflow for generating data stories, a crucial challenge remains: full automation is contingent upon the availability of predefined constraint sets, \ie, the \acp{dsl} for constraint representation. However, the origin of these \acp{dsl} poses a problem, as they are not readily available like off-the-shelf programming languages. In current practices, most \acp{dsl} are manually designed through the collaborative efforts of computer scientists and domain experts, a process that is both time-consuming and costly. This may be acceptable for specific applications requiring only a single \ac{dsl} library, as \ac{dsl} design is a \emph{once-and-for-all} endeavor there. Unfortunately, \acp{dsl} for representing constraints in data stories span multiple categories, diverse requirements, and an ever-expanding range of domains. For instance, within the ten constraint categories, there exist a vast array of potential \ac{dsl} instances. For the audience-oriented constraint, the \ac{dsl} syntax must be tailored to one specific audience group; for the domain-specific knowledge constraint, the \ac{dsl} semantics must encode the background knowledge of a particular domain of expertise; and let alone the narration constraint, the grammar of the \ac{dsl} must be designed on-the-fly based on the available evidence. Although it is conceivable that we derive a comprehensive set of constraints covering all potential domains, namely the so-called ``one-size-fits-all general constraint'', such an endeavor would result in a constraint system of prohibitively complexity, rendering it intractable for both machine and human end-users.

The highly varied and frequently evolving demands for \acp{dsl} are difficult to meet through human effort alone. Even if we manage to manually craft these \acp{dsl}, the progress in automated data story generation would be undermined --- we would merely be shifting human labor from one part of the workflow to another, even potentially increasing the overall labor required. Consequently, we find ourselves in a dilemma: \acp{gpl}, which easily accessible, are unsuitable for representing constraints due to their overwhelming complexity, whereas \acp{dsl}, which simplify specialized language features, inherently lack generalizability across different domains. To address this dilemma, rather than waiting for a universally applicable constraint to emerge, a more practical solution might involve automating the design of \ac{dsl}-based constraints. 

This solution is both feasible and evaluable. By adopting the AutoDSL approach~\cite{shi2024autodsl}, which combines bottom-up data-driven approaches and top-down principle-derived methods, we can automatically create \ac{dsl}-based constraints for data-driven storytelling based on relevant materials and design principles~\cite{shi2024abstract}. The resulting \acp{dsl} can be evaluated both quantitatively and qualitatively~\cite{fowler2010domain}. Quantitative evaluation checks the mapping from ontology elements in the reference model, \ie, concepts and relations in the domain corpus, to \ac{dsl} constructs of constraints; while qualitative evaluation takes the design guidelines of \ac{dsl} as questions for assessing the \ac{dsl}-based constraints, from a user-centric perspective. This joint pipeline of design and evaluation leads to a promising future where \ac{dsl}-based constraints are designed automatically, \ac{aigc} tools produce the necessary content and assets for data stories, and the constraints are synthesized and verified automatically. The integration of these approaches will enable content creators to script and implement their data stories more seamlessly.

\subsection{Valuing humans in data-driven storytelling}

Concerns may arise regarding the fully automation of data story generation and its potential severe impact on the content creation ecosystem. It appears that integrating constraints with generators could bridge the gap between creativity and preciseness, potentially marginalizing content creators. However, humans remain indispensable even in a future where constraints are fully realized. Firstly, constraints are not generators. While constraints define a feasible space for generation, generators determine which specific points within the space are sampled as the generated content. This indicates that the output space of \ac{aigc} tools remains significantly larger than the ideal output space that aligns with content creators on latent dimensions, such as aesthetic and ideological considerations. This disparity underscores the necessity for human-machine collaboration tools~\cite{li2024we}. Moreover, constraints can be latent. Even with automated \ac{dsl} design tools, not all constraints can be specified purely based on domain corpora. Some constraints require tacit knowledge from domain experts, necessitating a human-in-the-loop approach. Lastly, neither generators nor constraints can substitute for higher-level cognitive processes involving human factors, such as comprehensing, interpreting, and evolving the \emph{meaning} of data stories for humanity, and consequently, the metaphysical planning of seed ideas. Indeed, \ac{aigc} tools with constraints may merely alleviate human content creators from elementary technical tasks, thereby allowing them to concentrate on intention alignment, knowledge externalization, and metaphysical thinking.  

\acknowledgments{This work has been partially supported by RGC GRF Grant 16210722. The authors wish to thank Leixian Shen, Liwenhan Xie, Xian Xu, Hanlu Ma, Yanna Lin, Zhonghua Sheng, Shuchang Xu, Yuying Tang, Lin Gao, and Leni Yang for helpful discussions.}

\bibliographystyle{abbrv-doi}
\bibliography{references}
\end{document}